\documentclass[11pt]{amsart}

\usepackage{latexsym}
\usepackage{amsfonts}
\usepackage{amsthm}
\usepackage{graphicx}
\usepackage{amssymb}
\usepackage{amsmath}
\usepackage{amssymb}
\usepackage{color}
\author[]{Dima Grigoriev}
\address{Institut de Recherche Math\'ematique, Campus de Beaulieu, 35042 Rennes, France}
\email{dmitry.grigoryev@univ-rennes1.fr}

\author[]{Vladimir Shpilrain}
\address{Department of Mathematics, The City  College  of New York, New York,
NY 10031} \email{shpil@groups.sci.ccny.cuny.edu}
\thanks{Research of the second author was partially supported by
the NSF grant DMS-0405105. }

\newtheorem{definition}{Definition}
\newtheorem{proposition}{Proposition}
\newtheorem{remark}{Remark}

\def\Z{{\mathbf Z}}

\begin{document}

\title[]{Zero-knowledge authentication schemes\\
 from actions on graphs, groups, or rings}

\begin{abstract}
We propose a general way of constructing zero-knowledge
authentication schemes from actions of a semigroup on a set, without
exploiting any specific algebraic properties of the set acted upon.
Then we give several concrete realizations of this general idea, and
in particular, we describe several zero-knowledge authentication
schemes where  forgery (a.k.a.  impersonation) is NP-hard.
Computationally hard problems that can be employed in these
realizations include (Sub)graph Isomorphism, Graph Colorability,
Diophantine Problem, and many others.

\end{abstract}

\maketitle

\section{Introduction}

In this paper, we propose a general Feige-Fiat-Shamir-like
\cite{Fiat} construction of a zero-knowledge authentication scheme
from arbitrary actions.

Suppose a (partial) semigroup $S$ acts on a set $X$, i.e., for $s, t
\in S$ and $x \in X$,  one has $(st)(x)=s(t(x))$ whenever both sides
are defined.  For cryptographic purposes, it is good to have an
action which is ``hard-to-invert". We deliberately avoid using the
``one-way function" terminology here because we do not want to be
distracted by formal definitions that are outside of the main focus
of this paper. For a rigorous definition of a one-way function, we
just refer to one of the well-established sources, such as \cite{G}.
It is sufficient for our purposes to use an intuitive idea of a
hard-to-invert action which is as follows. Let $X$ and $Y$ be two
sets such that complexity $|u|$ is defined for all elements $u$ of
either set. A function $f: X \to Y$ is hard-to-invert if computing
$f(x)$ takes time polynomial in $|x|$ for any $x \in X$ (which
implies, in particular, that complexity of $f(x)$ is bounded by a
polynomial function of $|x|$), but there is no known algorithm that
would compute some $f^{-1}(y)$ in polynomial time in $|y|$ for every
$y \in f(X)$.

In our context of actions, we typically consider hard-to-invert
functions of the type $f_x : s \to s(x)$; in particular, a secret is
usually a {\it mapping}, which makes our approach different from
what was considered before. This idea allows us to construct a
general Feige-Fiat-Shamir-like zero-knowledge authentication scheme
from arbitrary actions, see the next Section \ref{protocol}. Then,
in the subsequent sections, we give several concrete realizations of
this general idea, and in particular, we describe several
zero-knowledge authentication schemes where recovering the prover's
secret key from her public key is an NP-hard problem. We note
however that what really matters for cryptographic security is
computational intractability of a problem on a {\it generic} set of
inputs, i.e., the problem should be hard on ``most" randomly
selected inputs. For a precise definition of the ``generic-NP"
class, we refer to \cite{ourbook}. Here we just say that some of the
problems that we employ in the present paper, e.g. Graph
Colorability, are {\it likely} to be generically NP-hard, which
makes them quite attractive for cryptographic applications.

We also address an apparently easier task of {\it forgery} (a.k.a.
{\it misrepresentation}, a.k.a. {\it  impersonation}), and show that
in most of our schemes this, too, is  equivalent for the adversary
to solving an NP-hard problem. To be more specific, by {\it forgery}
we mean the scenario where the adversary enters the authentication
process at the {\it  commitment} step, and then has to respond to
the {\it   challenge} properly.

Finally, we note that there were other attempts at constructing
zero-knowledge authentication schemes based on NP-hard problems
(e.g. \cite{Caballero1}, \cite{Caballero2}), but these constructions
are less transparent, and it is not immediately clear how and why
they work.


\section{Two protocols}
\label{protocol}

In this section, we give a description of two generic zero-knowledge
authentication protocols. Here Alice is the prover and Bob the
verifier.

\subsection{Protocol I}
Suppose a set $S$ acts on a set $X$, i.e., for any $s \in S$ and $x
\in X$, the element $s(x) \in X$ is well-defined.

\begin{enumerate}
\item Alice's public key consists of sets $S$, $X$,  an element $x \in X$, and an element  $u=s(x)$ for
some randomly selected $s \in S$, which is her private key.

    \item To begin authentication,
Alice selects an element $t \in S$ and sends the element $v=t(s(x))
\in X$, called the {\em commitment}, to Bob.
    \item
Bob chooses a random bit $c$, called the {\em challenge},  and sends
it to Alice.
\begin{itemize}
    \item
If $c=0$, then Alice sends the element  $t$  to Bob, and Bob checks
if the equality $v = t(u)$ is satisfied. If it is, then Bob accepts
the authentication.
    \item
If $c=1$, then Alice sends the composition  $ts$ to Bob, and Bob
checks if the equality $v = ts(x)$ is satisfied. If it is, then Bob
accepts the authentication.
\end{itemize}
\end{enumerate}

\subsection{Protocol II}
In this protocol, the hardness of obtaining the ``permanent" private
key for the adversary can be based on ``most any" search problem; we
give some concrete examples in the following sections, whereas in
this section, we give a generic protocol.

\begin{enumerate}
\item Alice's public key consists of a set $S$ that has a property ${\mathcal
P}$. Her private key is a {\it proof} (or a ``witness") that $S$
does have this property. We are also assuming that the  property
${\mathcal P}$ is preserved by {\it isomorphisms}.

\item To begin authentication, Alice selects an isomorphism $\varphi:
S \to S_1$ and sends the set $S_1$ (the commitment) to Bob.

\item
Bob chooses a random bit $c$ and sends it to Alice.
\begin{itemize}
    \item
If $c=0$, then Alice sends the isomorphism $\varphi$ to Bob, and Bob
checks (i) if  $\varphi(S) = S_1$ and  (ii) if  $\varphi$ is an
isomorphism.

    \item
If $c=1$, then Alice sends a proof of the fact that $S_1$  has the
property ${\mathcal P}$ to Bob, and Bob checks its validity.
\end{itemize}
\end{enumerate}

The following proposition says that in the Protocol II, successful
forgery is equivalent for the adversary to finding Alice's private
key from her public key, which is equivalent, in turn, to giving a
proof (or a ``witness") that $S$ does have the property ${\mathcal
P}$. The latter problem can be selected from a large pool of NP-hard
problems (see e.g. \cite{GJ}).

\begin{proposition}\label{general}
Suppose that after several runs of steps (2)-(3) of the above
Protocol II, both values of $c$ are encountered. Then successful
forgery in such a protocol is equivalent to finding a proof of the
fact that $S$ has the property ${\mathcal P}$.
\end{proposition}

\begin{proof}
Suppose Eve wants to  impersonate Alice. To that effect, she
interferes with the commitment step by sending her own commitment
$S_1'$ to Bob. Since she should be prepared to respond to the
challenge $c=0$, she should know an isomorphism $\varphi': S\to
S_1'$. On the other hand, since she should be prepared for the
challenge $c=1$, she should know a proof of the fact that $S_1'$ has
the property ${\mathcal P}$. Therefore, since $\varphi'$ is
invertible, this implies that she can produce a proof of the fact
that $S$  has the property ${\mathcal P}$.
 This completes the proof in one direction.

The other direction is trivial.
\end{proof}

\begin{remark} We note that finding a proof of the
fact that a given $S$ has a property ${\mathcal P}$ is not a
decision problem, but rather  a search problem (sometimes also
called a  {\em promise} problem), so we cannot formally allocate it
to one of the established complexity classes. However, we observe
that, if there were an  algorithm ${\mathcal A}$ that would produce,
for any $S$ having a property ${\mathcal P}$, a proof of that fact
in time bounded by a polynomial $P(|S|)$ in the ``size" $|S|$ of
$S$, then, given an arbitrary $S'$, we could run the algorithm
${\mathcal A}$ on $S'$, and if it would not produce a proof of $S'$
having the property ${\mathcal P}$ after running over the time
$P(|S'|)$, we could conclude that $S'$ does not have the property
${\mathcal P}$, thereby solving the corresponding decision problem
in polynomial time.

\end{remark}

\section{Graph isomorphism}
\label{graphiso}

In this section, we describe a realization of the Protocol I from
Section \ref{protocol} (actually, it also fits in with the Protocol
II), based on the Graph Isomorphism problem. We note that this {\it
decision} problem is in the class NP, but it is not known to be
NP-hard. Moreover, generic instances of this problem are easy,
because two random graphs are typically non-isomorphic for trivial
reasons. However, the problem that we actually use in the protocol
below, is a {\it promise} problem: given two isomorphic graphs, find
a particular isomorphism between them. This is not a decision
problem; therefore, if we want to allocate it to one of the
established complexity classes, we need some kind of
``stratification" to convert it to a decision problem. This can be
done as follows. Any isomorphism of a graph $\Gamma$ on  $n$
vertices can be identified with a permutation of the tuple $(1, 2,
\dots, n)$, i.e., with an element of the symmetric group $S_n$. If
we choose a set of generators $\{g_i\}$ of $S_n$, we can ask whether
or not there is an isomorphism between two given graphs $\Gamma$ and
$\Gamma_1$, which can be represented as a product of at most $k$
generators $g_i$. To the best of our knowledge, the question of
NP-hardness of this problem has not been addressed in the
literature, but it looks like a really interesting and important
problem.

\begin{enumerate}
\item Alice's public key consists of two isomorphic graphs, $\Gamma$
and $\Gamma_1$, and her private key is an isomorphism $\varphi:
\Gamma \to \Gamma_1$.

\item To begin authentication, Alice selects an isomorphism $\psi: \Gamma_1 \to
\Gamma_2$, and sends the graph $\Gamma_2$ (the commitment) to Bob.

\item
Bob chooses a random bit $c$ and sends it to Alice.
\begin{itemize}
    \item
If $c=0$, then Alice sends the isomorphism $\psi$ to Bob, and Bob
checks if $\psi(\Gamma_1) = \Gamma_2$ and if $\psi$ is  an
isomorphism.

    \item
If $c=1$, then Alice sends the composition  $\psi \varphi =
\psi(\varphi)$ to Bob, and Bob checks if $\psi \varphi(\Gamma) =
\Gamma_2$ and if $\psi  \varphi$ is  an isomorphism.
\end{itemize}
\end{enumerate}

A couple of comments are in order.

\begin{itemize}

\item As it is usual with Feige-Fiat-Shamir-like authentication
protocols, steps (2)-(3) of this protocol have to be iterated
several times to prevent a successful forgery with non-negligible
probability.

\item When we say that Alice ``sends"  (or  ``publishes") a graph, that
means that Alice sends  or publishes its adjacency matrix. Thus, the
size of Alice's public key is $2n^2$, where  $n$ is the number of
vertices in $\Gamma$.

\item When we say that Alice sends an isomorphism, that
means that Alice sends a permutation of the tuple $(1, 2, \dots,
n)$, where  $n$ is the number of vertices in the graph in question.
Thus, the size of Alice's private key is approximately $n\log n$.

\item When we say that Alice ``selects an isomorphism", that
means that Alice selects a random permutation from the group $S_n$;
there is extensive literature on how to do this efficiently, see
e.g. \cite{s}.

\end{itemize}

\begin{proposition}\label{prop_iso}
Suppose that after several runs of steps (2)-(3) of the above
protocol, both values of $c$ are encountered. Then successful
forgery in such a protocol  is equivalent to finding an isomorphism
between $\Gamma$ and $\Gamma_1$.
\end{proposition}

\begin{proof}
Suppose Eve wants to  impersonate Alice. To that effect, she
interferes with the commitment step by sending her own commitment
$\Gamma_2'$  to Bob.  Since she should be prepared to respond to the
challenge $c=0$, she should know an isomorphism $\psi'$ between
$\Gamma$ and  $\Gamma_2'$. On the other hand, since she should be
prepared for the challenge $c=1$, she should be able to produce the
composition  $\psi' \varphi = \psi'(\varphi)$. Since she knows
$\psi'$ and since $\psi'$ is invertible, this implies that she can
produce $\varphi$. This completes the proof in one direction.

The other direction is trivial.
\end{proof}

\section{Subgraph isomorphism}
\label{subgraphiso}

In this section, we describe another realization of the Protocol I
from Section \ref{protocol}, based on the Subgraph Isomorphism
problem. It is very similar to the Graph Isomorphism problem, but it
is {\it known} to be NP-hard, see e.g. \cite[Problem GT48]{GJ}. We
also note that this problem contains many other problems about
graphs, including the Hamiltonian Circuit problem, as special cases.
The problem is: given two graphs $\Gamma_1$ and $\Gamma_2$, find out
whether or not $\Gamma_1$ is isomorphic to a subgraph of $\Gamma_2$.
The relevant authentication protocol is similar to that in Section
\ref{graphiso}.

\begin{enumerate}
\item Alice's public key consists of two graphs, $\Gamma$
and $\Lambda_1$. Alice's private key is a subgraph $\Gamma_1$ of
$\Lambda_1$ and an isomorphism $\varphi: \Gamma \to \Gamma_1$.

\item To begin authentication, Alice selects an isomorphism $\psi: \Lambda_1 \to
\Gamma_2$, then embeds $\Gamma_2$ into a bigger graph $\Lambda_2$,
and sends the graph $\Lambda_2$ (the commitment) to Bob.

\item Bob chooses a random bit $c$ and sends it to Alice.
\begin{itemize}
    \item
If $c=0$, then Alice sends the subgraph  $\Gamma_2$ and the
isomorphism $\psi$ to Bob, and Bob checks if $\psi(\Lambda_1) =
\Gamma_2$ and if $\psi$ is  an isomorphism.

    \item
If $c=1$, then Alice sends the subgraph  $\Gamma_2$ and the
composition $\psi \varphi = \psi(\varphi)$ to Bob, and Bob checks
whether $\psi \varphi(\Gamma) = \Gamma_2$ and whether $\psi \varphi$
is  an isomorphism.
\end{itemize}
\end{enumerate}

Again, a couple of comments are in order.

\begin{itemize}
\item The Subgraph Isomorphism problem is NP-complete, see e.g.
\cite{GJ}.

\item When we say that Alice ``sends a subgraph" of a bigger graph, that
means that Alice sends the numbers $\{m_1, m_2, \dots, m_n\}$  of
vertices that define this subgraph in the bigger graph. When she
sends such a subgraph together with an isomorphism from another
(sub)graph, she sends a map  $(k_1, k_2, \dots, k_n) \to (m_1, m_2,
\dots, m_n)$ between the vertices.

\item Compared to the protocol in Section \ref{graphiso}, the size
of Alice's public key is somewhat bigger because Alice has to embed
one of the isomorphic graphs into a bigger graph. The size of
Alice's private key is about the same as in the protocol of Section
\ref{graphiso}.

\end{itemize}

\section{Graph colorability}
\label{color}

Graph colorability (more precisely, $k$-colorability) appears as
problem [GT4] on the list of NP-complete problems in \cite{GJ}. We
include an authentication protocol based on this problem here as a
special case of the Protocol II from Section \ref{protocol}. We note
that a (rather peculiar) variant of this problem was shown to be
NP-hard {\it on average} in \cite{Levin} (the latter paper deals
with edge coloring though).

\begin{enumerate}
\item Alice's public key is a $k$-colorable graph $\Gamma$, and her private key
is  a $k$-coloring of $\Gamma$, for some (public) $k$.

\item To begin authentication, Alice selects   an isomorphism  $\psi: \Gamma \to
\Gamma_1$, and sends the graph $\Gamma_1$ (the commitment) to Bob.

\item Bob chooses a random bit $c$ and sends it to Alice.
\begin{itemize}
    \item
If $c=0$, then Alice sends the isomorphism $\psi$ to Bob. Bob
verifies that $\psi$ is, indeed, an isomorphism from $\Gamma$ onto
$\Gamma_1$.

\item If $c=1$, then Alice sends a $k$-coloring of $\Gamma_1$ to Bob. Bob
verifies that this is, indeed, a  $k$-coloring of   $\Gamma_1$.

\end{itemize}
\end{enumerate}

Again, a couple of comments are in order.

\begin{itemize}

\item It is obvious that if $\Gamma$ is $k$-colorable and
$\Gamma_1$ is isomorphic to $\Gamma$, then $\Gamma_1$ is
$k$-colorable, too.

\item When we say that Alice ``sends a $k$-coloring", that
means that Alice sends a set of pairs $(v_i, n_i)$,       where
$v_i$ is a vertex and $n_i$ are integers between 1  and $k$ such
that, if $v_i$ is adjacent to $v_j$, then $n_i \ne n_j$.

\item Alice's algorithm for creating her public key (i.e., a $k$-colorable graph
$\Gamma$) is as follows. First she selects a number $n$ of vertices;
then she partitions $n$ into a sum of $k$ positive integers: $n=n_1+
\ldots + n_k$. Now the vertex set $V$ of the  graph $\Gamma$ will be
the union of the sets $V_i$ of cardinality $n_i$. No two vertices
that belong to the same $V_i$ will be adjacent, and any two vertices
that belong to different $V_i$ will be adjacent with probability
$\frac{1}{2}$. The $k$-coloring of $\Gamma$) is then obvious: all
vertices in the set $V_i$ are colored in color $i$.

\end{itemize}

\begin{proposition}\label{prop_color}
Suppose that after several runs of steps (2)-(3) of the above
protocol, both values of $c$ are encountered. Then successful
forgery is equivalent to finding a  $k$-coloring of   $\Gamma$.
\end{proposition}

The proof is almost exactly the same as that of Proposition
\ref{prop_iso}.

\section{Endomorphisms of groups or rings}
\label{endo_groups}

In this section, we describe a realization of the Protocol II (it
also fits in with the Protocol I) from Section \ref{protocol} based
on an algebraic problem known as the {\it endomorphism problem},
which can be formulated as follows. Given a group (or a semigroup,
or a ring, or whatever) $G$ and two elements $g, h \in G$, find out
whether or not there is an endomorphism of $G$ (i.e., a homomorphism
of $G$ into itself) that takes $g$ to $h$.

For some particular groups (and rings), the endomorphism problem is
known to be equivalent to the Diophantine problem (see \cite{Rom1,
Rom2}), and therefore the decision problem in these groups is
algorithmically unsolvable, which implies that  the related search
problem does not admit a solution in time bounded by any recursive
function of the size of an input.

Below we give a description of the authentication protocol based on
the endomorphism problem, without specifying a platform group (or a
ring), and then discuss possible platforms.

\begin{enumerate}
\item Alice's public key consists of a group (or a
ring) $G$ and two elements $g, h \in G$, such that $\varphi(g)=h$
for some endomorphism $\varphi \in End(G)$. This  $\varphi$ is
Alice's private key.

\item To begin authentication, Alice selects an automorphism $\psi$
of   $G$ and sends the element $v=\psi(h)$ (the commitment) to Bob.

\begin{itemize}
    \item
If $c=0$, then Alice sends the automorphism  $\psi$ to Bob, and Bob
checks whether  $v=\psi(h)$ and whether $\psi$ is  an automorphism.

    \item
If $c=1$, then Alice sends the composite endomorphism  $\psi \varphi
= \psi(\varphi)$ to Bob, and Bob checks whether  $\psi \varphi(g) =
v$ and whether $\psi \varphi$ is  an endomorphism.
\end{itemize}

\end{enumerate}

Here we point out that checking whether a given map is an
endomorphism (or an automorphism) depends on how the platform group
$G$ is given. If, for example, $G$ is given by generators and
defining relators, then checking whether a given map is an
endomorphism of $G$ amounts to checking whether every defining
relator is taken by this map to an element equal to 1 in $G$. Thus,
the {\it word problem} in $G$ (see e.g. \cite{LS} or \cite{ourbook})
has to be efficiently solvable.

Checking whether a given map is an automorphism is more complex, and
there is no general recipe for doing that, although for a particular
platform group that we describe in subsection \ref{metabelian} this
can be done very efficiently. In general, it would make sense for
Alice to supply a proof (at the response step) that her $\psi$ is an
automorphism; this proof would then depend on an algorithm Alice
used to produce $\psi$.

\begin{proposition}\label{prop_endomorph}
Suppose that after several runs of steps (2)-(3) of the above
protocol, both values of $c$ are encountered. Then successful
forgery is equivalent to finding an endomorphism $\varphi$ such that
$\varphi(g)=h$, and is therefore NP-hard in some groups (and rings)
$G$.
\end{proposition}

Again, the proof is almost exactly the same as that of Proposition
\ref{prop_iso}. We also note that in \cite{Grigoriev}, a class of
rings is designed for which the problem of existence of an
endomorphism between  two given rings from this class is NP-hard.

A particular example of a group with the NP-hard endomorphism
problem is given in the following subsection.

\subsection{Platform: free metabelian group of rank 2} \label{metabelian} A group $G$ is called {\it abelian}
(or commutative)  if $[a, b] =1$ for any $a, b \in G$, where $[a,
b]$ is the notation for $a^{-1} b^{-1} ab$. This can be generalized
in different ways. A group $G$ is called {\it metabelian} if $[[x,
y], [z,t]] =1$ for any $x, y, z, t \in G$.  The commutator subgroup
of $G$ is the group $G\,'=[G,G]$ generated by all commutators, i.e.,
by expressions of the form $[u,v] = u^{-1}v^{-1}uv$, where $u, v \in
G$. The second commutator subgroup  $G''$ is the commutator of the
commutator of $G$.

\begin{definition}
{\em Let $F_n$ be the free group of rank $n$. The factor group
$F_n/F_n''$ is called the {\em free metabelian group} of rank $n$,
which we denote  by $M_n$.}
\end{definition}

Roman'kov \cite{Rom2} showed that, given any Diophantine equation
$E$, one can efficiently (in linear time in the ``length" of $E$)
construct a pair of elements $u, v$ of the group $M_2$, such that to
any solution of the equation $E$,  there corresponds an endomorphism
of $M_2$ that takes $u$ to $v$, and   vice versa. Therefore, there
are pairs of elements of $M_2$ for which the endomorphism problem is
NP-hard (see e.g. \cite[Problem AN8]{GJ}). Thus, if a free
metabelian group is used as the platform for the protocol in this
section, then, by Proposition \ref{prop_endomorph}, forgery in that
protocol is NP-hard.

\subsection{Platform: $\Z_p^*$} Here the platform group is $\Z_p^*$,
for a prime $p$. Then, since $\Z_{p-1}^*$ acts on $\Z_p^*$ by
automorphisms, via the exponentiation, this can be used as the
platform for the Protocol I. In this case,  forgery is equivalent to
solving the discrete logarithm problem, by Proposition
\ref{prop_endomorph}.

\bigskip

\noindent {\it Acknowledgement.} The first author is grateful to Max
Planck Institut f\"ur Mathematik, Bonn for its hospitality during
the work on this paper.


\baselineskip 11 pt


\begin{thebibliography}{ABC}




\bibitem{Caballero1}
P. Caballero-Gil and C. Hern\'andez-Goya, {\it Strong Solutions to
the Identification Problem}, in:  7th Annual International
Conference COCOON 2001,  Lecture Notes Comp. Sc. {\bf  2108} (2001),
257--262.

\bibitem{Caballero2}
P. Caballero-Gil and C. Hern\'andez-Goya, {\it  A Zero-Knowledge
Identification Scheme Based on an Average-Case NP-Complete Problem},
in:  Computer Network Security, MMM-ACNS 2003, St. Petersburg,
Russia. Lecture Notes Comp. Sc. {\bf 2776} (2003), 289--297.

\bibitem{Fiat}
U. Feige, A. Fiat and A. Shamir, {\it Zero knowledge proofs of
identity,} Journal of Cryptology {\bf  1} (1987),  77--94.

\bibitem{GJ}
M. Garey, J. Johnson, {\em Computers and Intractability, A Guide to
NP-Completeness}, W. H. Freeman, 1979.



\bibitem{G}
O. Goldreich, {\it Foundations of cryptography}, Cambridge
University Press, 2001.

\bibitem{Grigoriev}
D. Grigoriev, {\it  On the complexity of the ``wild'' matrix
problems, of the isomorphism of algebras and graphs}, Notes of
Scientific Seminars of LOMI {\bf 105} (1981), 10--17 (in Russian)
[English translation in J. Soviet Math. {\bf 22} (1983),
1285--1289].

\bibitem{LS}
R.~C. Lyndon,  P.~E. Schupp,  \emph{Combinatorial Group Theory},
Ergebnisse  der Mathematik, band 89, Springer 1977.  Reprinted in
the Springer Classics in Mathematics series, 2000.

\bibitem{ourbook}
A. G. Myasnikov, V. Shpilrain, and A. Ushakov, {\it Group-based
cryptography}, Birkh\"auser, to appear.

\bibitem{Rom1}
V. A. Roman'kov, {\it Unsolvability of the problem of endomorphic
reducibility in free nilpotent groups and in free rings}, Algebra
and Logic {\bf 16} (1977),  310-–320.

\bibitem{Rom2}
V. A. Roman'kov, {\it Equations in free metabelian groups}, Siberian
Math. J. {\bf 20} (1979),  469-–471.

\bibitem{s}
A. Seress, {\it Permutation Group Algorithms}, Cambridge University
Press, 2002.

\bibitem{Levin}
R. Venkatesan, L. Levin, {\it  Random Instances of a Graph Coloring
Problem are Hard}, Proceedings of the Annual ACM Symposium on Theory
of Computing (1988), 217--222.



\end{thebibliography}
\end{document}